\documentstyle[aps,rotate,multicol,epsf]{revtex}
\def\beginwide{
        \end{multicols} \vspace*{-0.5cm} \noindent
        \rule{3.5in}{.1mm}\rule{.1mm}{5mm} \widetext \medskip }
\def\beginwidetop{
        \end{multicols} \vspace*{-0.5cm} \noindent
        \widetext \medskip }
\def\endwide{
        \hspace*{3.35in}~\rule[-5mm]{.1mm}{5mm}\rule{3.5in}{.1mm}
        \begin{multicols}{2} \vspace*{-1.0cm} \noindent }
\def\endwidebottom{
        \begin{multicols}{2} \vspace*{-1.0cm} \noindent }

\newcommand{\beq}{\begin{equation}}
\newcommand{\eeq}{\end{equation}}
\newcommand{\bea}{\begin{eqnarray}}
\newcommand{\eea}{\end{eqnarray}}

\draft

\begin{document}
\title{
Scaling and the prediction of energy spectra in decaying hydrodynamic turbulence
}
\author{P. D. Ditlevsen, M. H. Jensen and  P. Olesen\footnote{emails: pditlev@gfy.ku.dk, mhjensen@nbi.dk, polesen@nbi.dk} }
\address{Niels Bohr Institute,
Blegdamsvej 17, DK-2100 Copenhagen {\O}, Denmark }
\date{\today}
\maketitle
\begin{abstract}
Few rigorous results are derived for fully developed turbulence. By 
applying the scaling properties of the Navier-Stokes equation we have derived a
relation for the energy spectrum valid 
for unforced or decaying isotropic turbulence. 
We find the existence of a scaling function $\psi$.
The energy spectrum
can at any time by a suitable rescaling be mapped onto this function. This indicates that
the initial (primordial) energy spectrum is in principle retained
in the energy spectrum observed at any later time, and the principle of permanence of
large eddies is derived. The result can be seen as a restoration
of the determinism of the Navier-Stokes equation in the mean.  
We compare our results with a windtunnel experiment and find good agreement.
\end{abstract}
\begin{multicols}{1}

\vskip0.3cm
The Navier-Stokes equation (NSE) for hydrodynamic flow has been known for many years, and several
interesting numerical results have been found. However, very few analytic statements have 
been derived from these equations \cite{MY,Frisch}.
In this Letter we consider decaying isotropic turbulence, i.e.
hydrodynamics without external forcing. 

One well known feature of the hydrodynamic equations is their invariance
under re-scaling when the coordinates are scaled by an
arbitrary quantity $l$. Any solution to the NSE can then
be mapped onto another solution with a corresponding change in the 
velocity, the time and the diffusion 
coefficient $\nu$. The same scaling argument applies to the energy density. 
In the following we shall
show that this leads to a general scaling behavior of the energy density
considered as a function of $k,t,\nu$. 

The result resembles the $k^4$ backscattering at integral scales obtained in
EDQNM closure calculations \cite{lesieur:1997}. 
In studies of
the decay of homogeneous and isotropic turbulence self-similarity in the shape
of the energy spectra are often assumed, expressed as the principle
of permanence of large eddies (PLE)\cite{Frisch}. In a recent interesting
phenomenological study it was proposed that for initial spectra
not as steep as $k^4$ there will be three ranges of self-similarity
where PLE only applies to the very large scales \cite{eyink:2000}.

We shall now derive the scaling properties of the energy density directly from the
scaling properties of the NSE. This has 
been done by one of the authors in the case where viscosity was
ignored \cite{po}. Here we shall generalize these results. Consider
the energy density in $D$ dimensions, given by
\bea
&&{\cal E}(k,t,\nu,K,L)= \nonumber \\ 
&&\frac{\Omega_D k^{D-1}}{(2\pi)^DL^D}\int_{1/K}^L d^Dx
d^Dy~e^{i{\bf k(x-y)}}\langle{\bf v(x},t){\bf v(y},t)\rangle,
\label{espect}
\eea
where $L$ and $K$ are infrared and ultraviolet cutoffs, respectively, and
$\Omega_D$ is the solid angle. We assume isotropy such that the energy density
only depends on the modulus
$k=|{\bf k}|$ of the wave vector. The total energy density is given by
\bea
E(t)&=&\int_{1/L}^K dk \, {\cal E}(k,t,\nu,L,K) \nonumber \\ 
&=&\frac{1}{2}~\int_{1/K}^L d^Dx\, 
\langle{\bf v(x},t)^2\rangle.
\eea
Using the invariance of the unforced NSE under the scaling
\beq
{\bf x}\rightarrow l{\bf x}, ~~t\rightarrow l^{1-h}t,~~{\bf v}\rightarrow
l^h{\bf v},~~\nu\rightarrow l^{1+h}\nu,
\label{scaling}
\eeq
where $h$ is an arbitrary parameter, the relation
\beq
{\cal E}(k/l,l^{1-h}t,l^{1+h}\nu,K/l,Ll)=l^{1+2h}{\cal E}(k,t,\nu,K,L)
\label{basic}
\eeq
trivially follows.

In the following we assume the cutoffs such that for the relevant
physical range we have $1/L\ll k\ll K$. We will therefore suppress the
explicit dependence of the energy density on the cutoffs. The viscosity
provides a physical cutoff for large $k$ while the cutoff $1/L$ is determined 
by the boundary conditions at the integral scale.

Let us simplify (\ref{basic}) by introducing the new function $\psi$ by
defining
\beq
{\cal E}(k,t,\nu)=k^q~\psi (k,t,\nu).
\eeq
In the above equations we introduced the quantity $q=-1-2h$, which will be used in the following. 
To get more information, consider the relation (\ref{basic}) as a function
of $l$, differentiate with respect to $l$, and put $l=1$ afterward. This
yields the differential equation
\beq
-\frac{\partial\psi}{\partial\ln k}+{\textstyle \frac{q+3}{2}}\frac{\partial\psi}{\partial\ln t}
+{\textstyle \frac{1-q}{2}}\frac{\partial\psi}{\partial\ln\nu}=0.
\eeq
The general solution of this equation can be written in different equivalent forms:
\begin{eqnarray}
\psi (k,t,\nu)= \nonumber \\
F_\gamma({\textstyle \frac{q+3}{2}}\ln k+\ln t,{\textstyle \frac{(q+3)+(1-q)\gamma}{2}}\ln k+\ln t+\gamma\ln\nu)\label{scalingsolution}
\end{eqnarray}
where $F_\gamma$ is an arbitrary function, and $\gamma\ne 0$ is an arbitrary constant. 
These solutions correspond to
\beq
{\cal E}(k,t,\nu)=k^q~ \psi_1 \left(k^{\frac{3+q}{2}}t,\nu 
t^{-\frac{1-q}{3+q}}\right),
\label{I}
\eeq
for $\gamma=-(q+3)/(1-q)$ or
\beq
{\cal E}(k,t,\nu)=k^q~ \psi_2 \left(k^{\frac{3+q}{2}}t,\nu k^2t\right),
\label{II}
\eeq
for $\gamma=1$.
This is the
main result of this Letter, the consequences of which we will describe in
the following. 

If we take $t=0$ as the initial time, then the initial spectrum is immediately 
obtained as a power $k^q$ as stated before
(assuming that $\psi\neq 0$ for $t=0$). 
If the cutoffs $L, K$ had been explicitely included, they would appear as additional arguments 
$kL$ and $k/K$ in (\ref{I}) and (\ref{II}). 

If we simplify by assuming a very high Reynolds number flow with
a developed inertial range, then 
diffusion can be 
ignored for $k$ and $t$ sufficiently small such that the energy spectrum should 
have the form
\beq
{\cal E}(k,t,\nu)\approx k^q ~\psi \left(k^{\frac{3+q}{2}}t\right).
\label{approx}
\eeq
Note that this corresponds to a solution (\ref{scalingsolution}) with $\gamma=0$.
It is clear that in general the spectrum remembers the initial spectrum
through the factor in front of $\psi$, the scaling variable $k^{(3+q)/2}t$
and the functional form of $\psi$. The quantity $k^{(3+q)/2}t$ 
is invariant under the rescaling (\ref{scaling}) with the power 
$h=-(1+q)/2$. 
We must stress here that (\ref{approx}) is
only valid as a 'local in spectral space' approximation. Integrating 
(\ref{approx}) to obtain the total energy lead, as will be elaborated later,
to an inconsistency.

Shiromizu \cite{shiromizu} has shown from renormalization group
arguments that (\ref{approx}) emerges after long time, due to the
existence of a fixed point. 

From the scaling argument the actual form of the function $\psi (x)$ cannot be found.
If we impose the boundary condition that the Kolmogorov
spectrum should appear for large $k$'s and/or $t$, we must have
\beq
\psi (x)\sim x^{-2\frac{5+3q}{3(3+q)}},~~x=k^{\frac{3+q}{2}}t,~~~{\rm (3D)},
\label{kolmo}
\eeq
for large values of $x$.
This leads to
\beq
{\cal E}\sim k^{-5/3}~t^{-2\frac{5+3q}{3(3+q)}}.
\label{K41}
\eeq 
For even larger values of $x$ diffusion would of 
course become important. 
Now from classical K41 \cite{kolmogorov:1941b} dimensional counting we have ${\cal E}\sim k^{-5/3}\epsilon^{2/3}$,
where $\epsilon=dE/dt$ is the mean energy dissipation. Comparing
with (\ref{K41}) we have $dE/dt\sim t^{-(5+3q)/(3+q)}$ which after integration gives
\beq
E\sim t^{-2(1+q)/(3+q)}\label{decay}.
\label{Esimt}
\eeq  
Note that for $q \leq -1$ and $\psi(0)> 0$ we get from (\ref{approx}) an  infrared divergency in the
total energy. This reflects itself in an unphysical growth of
the energy with time, so for $q \leq -1$ 
(\ref{Esimt}) must be modified by an explicit dependence on
the cutoff $1/L$. 

In 2D one expects the
energy density to behave like $k^{-3}$, which corresponds to
\beq
\psi (x)\sim x^{-2},~~~{\rm (2D)},
\eeq
for large values of $x$. This gives
\beq
{\cal E}\sim k^{-3}t^{-2}
\eeq
for all $q$ except $q=-1$, where ${\cal E}=k^{-3}\psi (t)$.

The interpretation of the scaling as giving the development of an initial
power spectrum is of limited interest in practice. In the following we 
shall therefore show
that it is possible to use the scaling to obtain simple relations
between the energy densities at different times, allowing any ``initial''
spectrum compatible with NS without forcing. Let us assume that the
spectrum has been measured (or is otherwise known) at a time $t_0\neq 0$.
From the scaling we then obtain
\beq
{\cal E}_0(k,\nu)\equiv {\cal
E}(k,t_0,\nu)=k^q~\psi_1\left(k^{\frac{3+q}{2}}~t_0,
\nu~t_0^{-\frac{1-q}{3+q}}\right).
\label{N}
\eeq
We do not assume any special form for the initial energy density 
${\cal E}_0$. From the point of view of the scaling argument, the
quantity $q$ is an arbitrary parameter. Later we shall explain
how to determine $q$ for a given spectrum.

The relation (\ref{N}) can be used to construct the 
$k-$dependence of the scaling
function from the measured ${\cal E}_0$ by a simple shift of variables,
\begin{eqnarray}
\psi_1\left(k^{\frac{3+q}{2}}~t,\nu~ t_0^{-\frac{1-q}{3+q}}\right)= \\
k^{-q}~\left(t/t_0\right)^
{-\frac{2q}{3+q}}~{\cal E}_0\left(k~\left(t/t_0\right)^{\frac{2}{3+q}},\nu\right).\nonumber
\end{eqnarray}
This in turn leads to the following relation between the energy densities at
different times,
\beq
{\cal E}\left(k,t,\nu~\left(\frac{t_0}{t}\right)^{\frac{1-q}{3+q}}\right)=\left(\frac{t_0}{t}\right)^{\frac{2q}{3+q}}~
{\cal E}_0\left(k~\left(\frac{t}{t_0}\right)^{\frac{2}{3+q}},\nu\right).
\label{N1}
\eeq
This relation is a rewriting of the basic equation (4).
It is an expression of what could be called {\it determinism in the mean} in
decaying turbulence, which indeed is not predictable in time.
We see that in spite of the chaotic behavior of the velocity, it is in 
principle possible to predict the energy density from
knowing it at a fixed time. This predictive power is strongest in a
region where diffusion can be ignored. Otherwise  (\ref{N1}) makes
predictions of the energy densities for different times and different
viscosities, and relevant experiments may be hard to carry out in practice. 

On the right hand side of 
 (\ref{N1}) the time dependence of the argument proportional to $\nu$ is
absent for $q=1$ and is 
very weak for $q$ not too far from 1. In the application to be discussed below
we see no effect of diffusion, except for large $k$.

The scaling arguments leading to the above results are based on a scaling
of the time by a positive quantity. Taking time itself to be positive, it 
follows that the earliest time we can have is $t=0$. We shall now discuss how
the parameter $q$ can be determined in principle. In doing this we consider
the high Reynolds number limit  
by taking $\nu\rightarrow 0$ (for $q=1$ this is not necessary). Then 
 (\ref{N1}) reduces to
\beq
{\cal E}(k,t,\nu \rightarrow 0)=(t_0/t)^{\frac{2q}{3+q}}~{\cal E}_0\left(k~(t/t_0)^
{\frac{2}{3+q}},\nu\rightarrow 0\right).
\label{N2}
\eeq
In the limit $t\rightarrow 0$ existence of the spectrum on the left
hand side requires
\beq
{\cal E}_0\left(k~(t/t_0)^{\frac{2}{3+q}},\nu \rightarrow 0\right)\sim \left(k(t/t_0)^
{\frac{2}{3+q}}\right)^{q}.
\label{N3}
\eeq
In  (\ref{N2}) this implies
\beq
{\cal E}(k,t,\nu \rightarrow 0)\sim k^q~~~{\rm for}~~t\rightarrow 0.
\eeq
Thus we see that irrespective of whether the spectrum is measured at the 
initial time, it must behave like $k^{q}$ at the earliest possible time, 
in the limit $\nu \rightarrow 0$, which means that $k$ should not be too large.
Even if the experiment was not operative at $t=0$, one can extrapolate
back to a hypothetical initial time $t=0$.

From (\ref{N3}) it also follows that
\beq
{\cal E}(k,t,\nu \rightarrow 0)\sim k^q~~{\rm for}~~k\rightarrow 0.
\eeq
Thus, for all times the spectrum keeps its original form for very low
values of $k$. This is the principle of permanence of large eddies.
It allows us to fix the parameter $q$ if experimental results exist for
sufficiently small $k$ by fitting the data to $k^q$. It should be noticed 
that we have only been able to 
prove the principle of permanency of large eddies when $\nu \rightarrow 0$, except for the case
$q=1$, where $\nu$ is not mixed up with the time in (\ref{N1}).  A
weaker assumption used in the proof is the existence of the spectrum
for $t\rightarrow 0$, which imposes the boundary condition (\ref{N3})
on ${\cal E}_0$.

In order to test these results consider the experimental measurements of the energy
spectrum in decaying turbulence by 
Comte-Bellot and Corrsin \cite{comte-bellot:1971}. The energy spectrum was
measured in a windtunnel experiment at different distances from the turbulence
generating grid,
see the figure (a) and (d)). 
The scaling parameter $q$ is determined from the data by using (\ref{decay}). 
It is found that $E(t) \sim t^{-1.4}$  which is in the range of most
experimental findings \cite{mohamed:1990}.
In panels (b) and (e) this is done 
and from the scaling we obtain in both cases $q=3.67$. 
The
panels (c) and (f) show the scaling function $\psi(k^{(3+q)/2}t)={\cal E}(k,t,\nu)/k^q$ 
obtained from   (\ref{approx}) as a function of $(k^{(3+q)/2} t)^{2/(3+q)}$. The collapse is
seen to be perfect. The energy spectra show an inertial range of less than two
decades, and even less for the 1 inch grid experiment. So even in this case of relatively
low Reynolds number flow the dependence of the scaling function on $\nu$ is weak and
(\ref{approx}) holds
very well.   
It should, furthermore, be noticed that the factor 
$(t/t_0)^{(1-q)/(3+q)}$, which multiplies $\nu$ in (\ref{N1}) for
these experiments vary from 1 to 0.42, i.e. by a factor less than
2.4. That could as well be the reason why the dependence on $\nu$ 
can not be detected in the experimental data.
  
In general, the energy cannot be obtained by integration over the energy density, 
unless one has some knowledge of the scaling 
function $\psi$. For example, without such knowledge
one cannot judge the convergence of the relevant integral. 
In the case $q=1$ the situation is, however, quite different, since  
(\ref{I}) shows that 
\beq
{\cal E} (k,t,\nu)=k~\psi_1\left(k^2t,\nu\right)\equiv k~\psi 
\left(k^2t\right),
\eeq
where we left out any explicit reference to the diffusion coefficient 
since it is the same on both sides.
In principle, this equation can be tested experimentally as a check of the
validity of the NSE.

Consider now the total energy, 
\beq
E(t)=\int_0^\infty dk ~k\psi (k^2t)= \frac{1}{2t}~\int_0^\infty dy ~\psi
 (y),
\eeq
which shows that in the case $q=1$ the NSE predict a $1/t$
decay of the total energy. Since diffusion is included the integral is
expected to converge rapidly. The inertial range, where the energy is
constant, would be defined by introducing a cutoff $y_0(t)$ on the
$y-$integral. Then one should require
\beq
\int_0^{y_0(t)} dy ~\psi (y)\propto t
\eeq 
The Kolmogorov scale $y_0(t)$ can of course only be determined from knowledge
of the function $\psi$. 
In order to obtain the energy in cases with $q\neq 1$, one would also
need some knowledge of the scaling function.
The classical problem of assuming a scaling relation of the form (\ref{approx})
where the diffusion is ignored all together is illuminated by considering 
the requirement of energy conservation.
This leads to the exact relation
\begin{eqnarray}
\frac{dE}{dt}&=&\frac{d}{dt}\int_0^\infty dk~{\cal E}(k,t)\nonumber \\
&=&-2\nu\int_0^\infty dk~k^2{\cal E}(k,t).
\end{eqnarray}
Using (\ref{I}) after a shift of integration variable this can be expressed as
\bea
&&{\textstyle \frac{2(1+q)}{(3+q)^2}}\int_0^\infty dx~x^{\frac{q-1}{3+q}}~\psi_1 (x,z) 
\label{constraint}\\
&+&
{\textstyle \frac{1-q}{3+q}}\int_0^\infty dx~
x^{\frac{q-1}{3+q}}~z\partial_z\psi_1 (x,z)
=2 z\int_0^\infty dx~x\psi_1 (x,z),\nonumber
\eea
where the variable $x$ is defined in (\ref{kolmo}) and $z=\nu~t^{-\frac{1-q}{3+q}}$
For $q=1$ this simplifies to
\beq
\int_0^\infty dx~\psi_1 (x,\nu)=2\nu\int_0^\infty dx~x\psi_1 (x,\nu).
\label{q=1}
\eeq
This relation gives a rather weak constraint on the scaling function.
It is now seen why the second argument in $\psi_1$ cannot be ignored:
For $q\neq 1$ 
(\ref{constraint}) cannot be satisfied, since this would
lead to the meaningless requirement $t^{(q-1)/(3+q)}=$constant. However,
when the argument $z$ is included there is no such problem, since all
terms in (\ref{constraint}) are $z-$dependent.

Eq. (\ref{q=1}) simply states that $\langle x\rangle=1/(2\nu)$. Thus, the mean value of the
scaling
variable is determined by the diffusion, and $\langle k^2\rangle=1/(2\nu t)$. Thus, for
large times, the mean value moves toward small $k$, as expected, with a rate
given by $\nu$. In a similar way we can analyze the content of   
(\ref{constraint}). 
Defining the quantity
\beq
\langle\langle x^\alpha\rangle\rangle =\int_0^\infty dx~x^\alpha~\psi_1 (x,z),
\label{distrib}
\eeq
then (\ref{constraint}) has the solution:
\beq
\langle\langle k^{(q-1)/2}\rangle\rangle /\langle\langle k^{(3+q)/2}\rangle\rangle =\mbox{const.} \nu t.
\eeq
This can be interpreted by introducing a weight function $w(k)=k^{(q-1)/2}$,
from which we obtain
\beq
\langle k^2\rangle_w \sim 1/\nu t,
\label{eddydiffusion}
\eeq
so $k$ approaches zero as time goes by, with a rate again given by $\nu$.
It is interesting to note that (\ref{eddydiffusion}) behaves exactly as a
spectral diffusion. 

To summarize, we have derived a rigorous scaling relation in the case of
decaying hydrodynamic turbulence with a given
initial state. This is expressed
in terms of the function $\psi$ which itself is not a power function, but
depends on the initial scaling. This means that the information of the primordial
state is still on average present at later times. The information travels
from the small scales to the large scales
(or backwards in $k$-space) in a diffusive way.

Useful comments from E. Aurell, G. Eyink and V. Yakhot are gratefully acknowledged.

\end{multicols}

\begin{figure}[ht]
\vbox{
{
        \centerline{\epsfxsize=17cm
        {\epsfbox{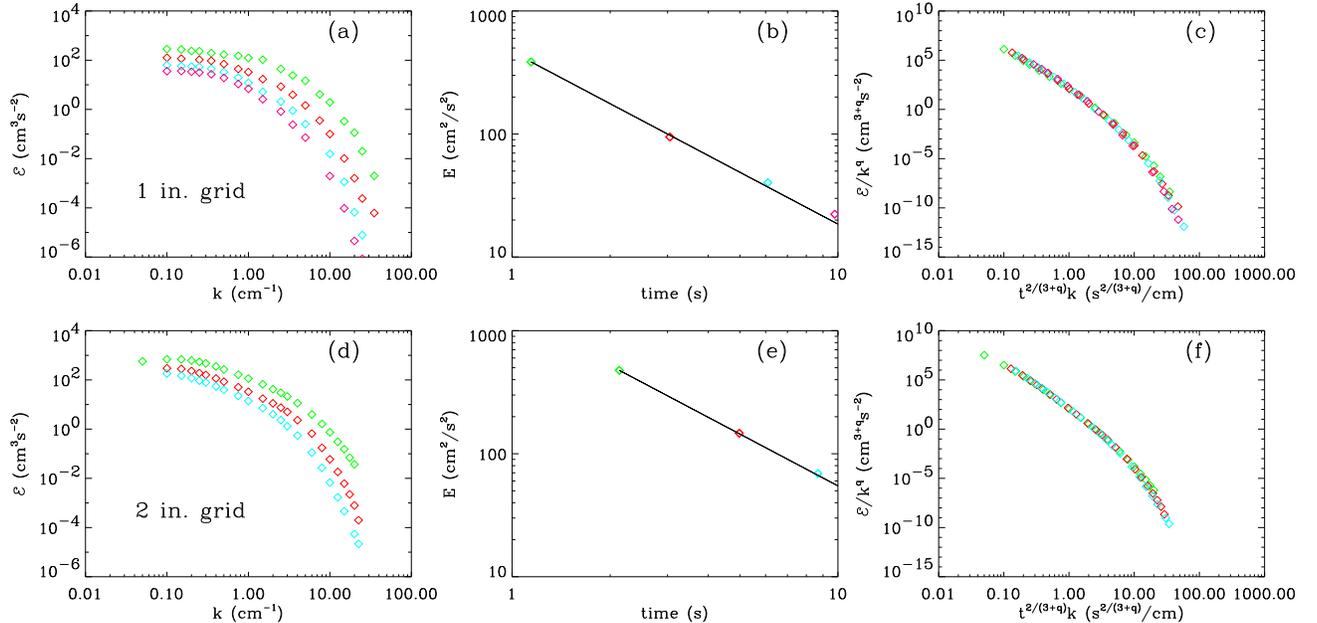}}}
}
}
\caption{
Decay of grid generated turbulence in a windtunnel (Comte-Bellot and Corrsin, 1971)
Panels (a) and (d) show energy spectra in two experiments with grid spacing of 1 inch and 2 inches respectively. The kinematic viscosity is that
of air, $\nu = 1.5 \times 10^{-5}~m^2/s$.
The spectra are measured at different distances from the grid, which by Taylors hypothesis
is converted into times using $t=D/U_0$, where $D$ is the distance from the mesh and $U_0$ is
the wind velocity upstream from the grid. The times are in panel (a): t=1.14 s (green), t=3.05 s 
(red), t=6.1 s (blue), t=9.8 s (purple), and in panel (d): t=1.07 s (green), t=2.49 s (red) and
t=4.34 s (blue). 
Panels (b) and (e) show the total energy $E(t)$ as a function of time $t$ with
the color coding as in panels (a) and (d). The straight lines are $E \sim t^{-2(1+q)/(3+q)}$ with
$q=3.67$. Panels (c) and (f) show the scaling function $\psi(k^{(3+q)/2}t)={\cal E}(k,t,\nu)/k^q$
as a function of $k t^{2/(3+q)}$ with $q=3.67$.
}
\end{figure}
\end{document}